\documentclass[final,5p,times,twocolumn]{elsarticle}

\usepackage{latexsym}
\usepackage{graphics}
\usepackage{graphicx}
\usepackage{dcolumn}
\usepackage{amsmath}
\usepackage{textcomp}
\usepackage{amssymb}
\usepackage{cite}
\usepackage{subfigure}
\usepackage[section]{placeins}
\usepackage{fix-cm}
\usepackage{color}

\newcommand{\figref}[1]{Figure \ref{#1}}
\newcommand{\eref}[1]{Eq.(\ref{#1})}
\newcommand{\erefs}[1]{Eqs.(\ref{#1})}
\newcommand{\reff}[1]{(\ref{#1})}
\newcommand{\p}{\partial}
\newcommand{\emphr}[1]{{#1}}
\newcommand{\rb}{\bar{r}}
\newcommand{\oso}{\omega_{\sigma_{\!\!\perp}}}
\newcommand{\BB}{\tilde{B}_1}
\newcommand{\nn}{\nonumber}
\newcommand{\q}{\quad}
\newcommand{\qq}{\qquad}

\journal{}

\begin{document}

\begin{frontmatter}



\title{Thermo-Galvanometric Instabilities in Magnetized Plasma Disks}


\author[a1]{Alessio Franco}
\author[a2,a1]{Giovanni Montani}
\author[a1]{Nakia Carlevaro}
\address[a1]{Physics Department, ``Sapienza'' University of Rome, Piazzale Aldo Moro 5, (00185) Roma, Italy.}
\address[a2]{ENEA - Unit\`a Tecnica Fusione, ENEA C.R. Frascati, Via E. Fermi 45, (00044) Frascati (Roma), Italy.}

\begin{abstract}
In this work, we present a linear stability analysis of fully-ionized rotating plasma disks with a temperature gradient and a sub-thermal background magnetic field (oriented towards the axial direction). We describe how the plasma reacts when galvanometric and thermo-magnetic phenomena, such as Hall and Nernst-Ettingshausen effects, are taken into account, and meridian perturbations of the plasma are considered. It is shown how, in the ideal case, this leads to a significant overlap of the Magneto-rotational Instability and the Thermo-magnetic one. Considering dissipative effects, an overall damping of the unstable modes, although not sufficient to fully suppress the instability, appears especially in the thermo-magnetic related branch of the curve.
\end{abstract}



\end{frontmatter}



\section{\textbf{Introduction}}
\label{intro}

The dynamical problem of matter falling on the surface of a central astrophysical object dates back to the work of Kuiper, in 1941 \citep{kuiper}. Long before the actual observation of the phenomenon, in that work was inferred that matter would have slowly spiraled inward, thus forming a rotating disk around the star. From there, accretion disk physics and dynamics have been thoroughly studied, mainly due to their connection to crucial phenomena such as Gamma Ray Bursts \citep{grb} and Active Galactic Nuclei \citep{agn}, as well as due to their peculiar observed luminosities, several orders of magnitude higher than those from the corresponding stars. In order to appreciate the relevance of this emission, we emphasize how a compact X-binary disk can emit up to $\mathcal{O}(10^{37})\;\textrm{erg}\;\textrm{s}^{-1}$, whereas the Sun shows a four-orders of magnitude lower luminosity. The problem is that such an efficient energy extraction and transport to the outer layers of the disks has no simple physical interpretation. The reason lies in the fact that the outflow of energy can not be explained by using the kinetic viscosity of the disk, but instead this phenomenon requires the plasma to be turbulent in order to trigger an effective viscosity.

It has been shown that a linear instability can be triggered in a pure hydrodynamical disk in differential rotation, but only if the vorticity of the plasma (angular momentum per unit of mass) decreases outwards \citep{BKL01}. However, this is a requirement generally not fulfilled by astrophysical systems. The idea that a background magnetic field could be an effective source of enhanced transport is one already presented by Lynden Bell \citep{lynden,lynden74} and Shakura and Sunyaev \citep{shakurasunyaev,shakura}, but it was with the works by Balbus and Hawley \citep{BH91,mri2,mri3,mri4} that the magnetic field could be directly linked to turbulence via the linear instability that it causes in the plasma flows. This is called Magneto-rotational Instability (MRI) and it holds for a differentially rotating thin disk as far as an arbitrarily small magnetic field is involved in the problem. Such an instability is nonetheless affected by specific restriction. First of all, MRI is suppressed for perturbation having sufficiently small wave-length. Moreover, assuming a vertical magnetic field, the disk is actually stable for perturbations propagating in the radial direction only. Finally, for the instability to take place, the following condition must be verified: $v_{A}^{2}>6v_{s}^{2}/\pi^{2}$, in which $v_{A}=B/\sqrt{4\pi\rho}$ is the Alfv\'en velocity of the plasma and $v_{s}=\sqrt{5P/3\rho}$ the (adiabatic) sound speed (here, $B$, $P$ and $\rho$ stand for the disk magnetic field, gas pressure and density, respectively): this means that the magnetic field has to be sub-thermal, \emph{i.e.}, in terms of the $\beta$ plasma parameter we get $\beta>1$ (we remind that for an adiabatic equation of state, it results $\beta\equiv8\pi P/B^{2}=6v_s^{2}/5v_A^{2}$). 

Since the stability emerges in the small $\beta$ and spacial scale region, we investigate if other collisional contribution to the generalized Ohm can induce new kind of instabilities in such regimes, \emphr{by also characterizing their absolute or convective nature \citep{LLFM,HM90}}. A similar approach, in the case of axially propagating perturbations, has already been carried out in \citep{urpin}, analyzing an analogous yet substantially different physical configuration. In fact, in our study, we consider also a non-zero thermal transport in the plasma, thus the energy equation has to be included in the dynamical system and we succeed in fixing a Thermo-magnetic Instability (TMI), generalizing the analysis in \citep{MNRAS}. Moreover, our calculation stands for all spatial scales, holds for radial perturbations too, and also for arbitrarily large values of the magnetic field, \emph{i.e.}, it does not depend on $\beta$ (note, however, that the choice of the plasma parameter actually affects the magnitude of the instability).

In particular, we analyze the case of local meridian perturbations of the disk parameters \emphr{by taking into account the galvanometric and thermo-magnetic contributions to the system, which corresponds to include the Hall and Nernst terms in the generalized Ohm law, respectively. In defining how exactly these modifications affect the dynamics of the instabilities, we start by considering, in a non-dissipative environment, the two effects separately} and then both of them along with the thermo-electromotive force the plasma is subject to. While the first case results to be not much different from the pure MRI, the second shows an overlap of the MRI \emphr{absolute} modes with the thermo-magnetic ones, thus extending the instability to all scales. \emphr{It is worth noting how the convective instability results still dominant for large spatial scales too. In fact, the complex and conjugated solutions outline a significant increase of the growth rate as the wave number decreases, finding its cutoff in the natural size of the disk as well as in the validity region of the considered local approximation}. We complete our analysis by considering all the above mentioned collisional effects, in the presence of dissipative terms (namely, the electrical and the thermal conductivities). We show how, in such a Thermo-galvanometric Instability (TGMI) scenario, the perturbed system is subjected to a general damping, thus stabilizing a considerable range of the allowed wave-vector values. We shall outline how MRI and TMI are somewhat complementary.

The paper is organized as follows. In Sec.\ref{sec:1}, we present the set of the basic equations which describe the local behavior of the perturbed system. \emphr{In Sec.\ref{sec:2}, after a brief description of the approximations used and of the main features of MRI and TMI, respectively, we focus on the different kinds of absolute instabilities generated from the presence of each thermo-magnetic effect, in the ideal case of meridian perturbations. A discussion on the emerging convective (propagative) instabilities is presented in the corresponding Subsection}. In Sec.\ref{sec:3}, the analysis on the role of the dissipative terms in the configuration is developed in the limit of large, as well as, small $\beta$ parameters. Concluding remarks follow.

\section{Basic Formalism}
\label{sec:1}

The fundamental equations of the accretion disk structure are basically those of MHD, but considering that Hall and Nernst effects modify the expressions of the electric field $\textbf{E}$ and the heat density flux vector $\textbf{q}$. This results in changing the frozen-in and energy conservation laws; hence, the heat density flux vector and the generalized Ohm law become
\begin{align}
\textbf{q}=\alpha_{\parallel}T\textbf{j}_{\parallel}+\aleph T\textbf{B}\times\textbf{j}-\chi_{\parallel}\left({\nabla}T\right)_{\parallel}-\chi_{\perp}\left({\nabla}T\right)_{\perp}\;,\label{eq:calore}\\
\textbf{E}+\frac{1}{c}\textbf{v}\times\textbf{B}=\frac{\textbf{j}_{\parallel}}{\sigma_{\parallel}}+\frac{\textbf{j}_{\perp}}{\sigma_{\perp}}+\Re\textbf{B}\times\textbf{j}+\qq\qq\nn\\
+\alpha_{\parallel}\left({\nabla}T\right)_{\parallel}+\aleph\textbf{B}\times{\nabla}T\;,\label{eq:ohmgeneralizzata}
\end{align}
respectively, where we have adopted the following notation: $\textbf{B}$ is the magnetic field (bold stands for vector while $B=|\textbf{B}|$), $\chi$ is the thermal conductivity, $\sigma$ is the electrical conductivity, $\textbf{v}$ is the plasma speed, $\textbf{j}$ is the plasma current density, $T$ is the plasma temperature; the subscripts $\parallel$ and $\perp$ indicate components parallel and perpendicular to the magnetic field lines (in cylindrical coordinates [$r$, $\phi$, $z$]). $\Re$, $\aleph$ and $\alpha$ stand for the Hall, Nernst and thermo-electromotive coefficients, respectively, and their approximate expression can be derived using kinetic theory \citep{LP81} to obtain
\begin{align}\label{coefs}
\Re=-\frac{1}{ecn}\;,\q
\aleph=-\frac{\nu_{ei}}{\sqrt{2\pi}\;mc\omega_{Be}^{2}}\;,\q
\alpha_{\parallel}=\frac{1}{e}\Big(\frac{\mu}{T}-4\Big)\;,
\end{align}
in which $e$, $m$ are the (positive) electron charge and mass, $c$ is the speed of light, $n$ and $k_{B}$ are the number density and the Boltzmann constant, respectively; the electron-ion collisional frequency $\nu_{ei}$ and the electron gyrofrequency $\omega_{Be}$ are defined, denoting with $L_{e}$ the Coulomb logarithm, by:
\begin{align}
\nu_{ei}=\frac{4\pi e^{4}L_{e}n}{m^{1/2}\,T^{3/2}}\;,\qquad\qquad
\omega_{Be}=\frac{eB}{mc}\;.
\end{align}
while the chemical potential $\mu$ is described by the following relation
\begin{equation}
\mu=T\ln\big[n h^{3}\left(2\pi mT\right)^{-3/2}\big]\;.
\end{equation}

It is important to stress that we have intentionally neglected the orthogonal term of the thermo-electromotive force in the Ohm law, \emph{i.e.}, $\alpha_{\bot}(\nabla T)_{\bot}$, since in \citep{MNRAS} it has been shown that such a term is negligible with respect to the Nernst contribution. In particular, the expression of $\alpha_\bot$ is \citep{LP81,BRA65} $\alpha_{\bot}\simeq0.36(\nu_{ie}/\omega_{Be})^{2}$, while the Nernst coefficient can be rewritten from \eref{coefs} as $\aleph\simeq10^{10}(\nu_{ie}/\omega_{Be})^{2}$ for typical parameter values of the accretion disk (here, we have set $T=10^{4}$K and $n=10^{9}$cm$^{-3}$). We recall that the adopted expressions for the kinetic coefficients are only valid in the limit of sufficiently strong magnetic fields, \emph{i.e.}, $\omega_{Be}\gg\nu_{ee}$ (here, $\nu_{ee}$ denotes the electron-electron collisional frequency).

\erefs{eq:ohmgeneralizzata} and \reff{eq:calore} have to be combined with Maxwell equations (Faraday and Ampere laws) and the conservation law of internal energy for a one-atom perfect gas \citep{jackson,ortolani}:
\begin{align}
\nabla\times\textbf{\textbf{E}}=-\p_{t}\textbf{\textbf{B}}/c\;,\\
\nabla\times\textbf{B}=4\pi\textbf{j}/c\;,\label{eq:maxwell}\\
\partial_{t}P=-\tfrac{5}{3}P\left({\nabla}\cdot\textbf{v}\right)-\tfrac{2}{3}{\nabla}\cdot\textbf{q}\label{eq:ort}\;,
\end{align}
respectively, note that displacement currents are taken here to be negligible since the fluid motion is non-relativistic. 

The analysis is implemented  at a fixed distance $r=\rb$ from the central body of mass $M$ and the accretion disk is assumed to be thin, \emph{i.e.}, the half-depth $H(r)$ verifies the inequality $H(\rb)\ll\rb$. The local equilibrium configuration is described by a first-order perturbation of the MHD equations described above and we denote the background with $(...)_0$ and the fluctuations with $(...)_1$. A generic variable $A$ is thus perturbed near $\rb$ as $A=A_0(\rb)+A_1(r)$, with $A_1\ll A_0$. It is important to stress that the effects associated with $\textbf{q}$ are neglected in the background dynamics being determined by gravity only. Moreover, we are dealing with small scale perturbations and the radial variation of the zeroth-order quantities can be effectively frozen to a given fiducial radius.

The zeroth-order configuration is determined for an adiabatic equation of state $P_0\sim\rho_0^{5/3}$ (where $\rho_0(\rb,z)$ is the disk mass density). The background radial momentum conservation (which reduces to the balance of the gravitational force with the centripetal one) locally fixes the Keplerian nature of the disk angular frequency as $\omega_k=\sqrt{GM/\rb^3}$. On the other hand, the zeroth-order vertical local equilibrium determines the gravostatic profile of decay for the mass density (and for the pressure) as the vertical coordinate increases. The equilibrium between the pressure and the vertical gravitational force simply results into the background profile \citep{BKL01} $\rho_0(\rb,z)=\tilde{\rho}[1-z^2/H(\rb)^2]^{3/2}$, where $H^2=2\tilde{P}/\tilde{\rho}\omega_k^2$ (pressure being $\tilde{P}\sim\tilde{\rho}^{5/3}$, with $\tilde{\rho}=const.$). In this work, the $z$-dependence is however disregarded in each background variables in view of the thinness of the disk ($|z|\leq H(\rb)\ll\rb$), thus reducing the mass density to $\rho_0=\tilde{\rho}$, the pressure to $P_0=\tilde{P}$ and the temperature to $T_0=\tilde{T}\equiv const.$, \emph{i.e.}, to assigned constants for the stability problem. While, the background magnetic and velocity fields read as 
\begin{align}
\textbf{B}_0=(0,\,0,\,B_{0z})\;,\qquad\quad \textbf{v}_0=(0,\,\rb\omega_k,\,0)\;,
\end{align}
respectively.

In this formalism, the perturbed magnetic induction equation reads
\begin{multline}
\partial_{t}\textbf{B}_{1}=-\textbf{B}_{0}{\nabla}\cdot\textbf{v}_{1}+\frac{c^{2}\Re}{4\pi}{\nabla}\times\left(\textbf{B}_{0}\cdot{\nabla}\right)\textbf{B}_{1}+\\
-c\alpha_{\parallel}{\nabla}\times\left({\nabla}T_{1}\right)_{\parallel}+c\aleph\left[\left(\textbf{B}_{0}\cdot{\nabla}\right){\nabla}T_{1}-\textbf{B}_{0}\nabla^{2}T_{1}\right]+\\
+\omega_{k}\left(B_{1r}\textbf{e}_{\phi}-B_{1\phi}\textbf{e}_{r}\right)\;,\label{eq:congelamento completo}
\end{multline}
here and in the following we indicate with $(\textbf{e}_{r},\,\textbf{e}_{\phi},\,\textbf{e}_{z})$ the cylindrical versors. While the (perturbed) continuity and Navier-Stokes equations \citep{mri} are
\begin{align}
\partial_{t}\rho_{1}+\rho_{0}{\nabla}\cdot\textbf{v}_{1}=0\;,\label{eq:Continuità}\\
\rho_{0}\left[\partial_{t}\textbf{v}_{1}+\left(\textbf{v}_{0}\cdot{\nabla}\right)\textbf{v}_{1}\right]=-{\nabla}P_{1}+\nn\qq\qq\qq\qq\\
-\frac{1}{4\pi}\left[{\nabla}\left(\textbf{B}_{0}\cdot\textbf{B}_{1}\right)-\left(\textbf{B}_{0}\cdot{\nabla}\right)\textbf{B}_{1}\right]\;,\label{eq:Conservazioneqdmvettoriale}
\end{align}
respectively. \eref{eq:ort} can be written as
\begin{multline}
\partial_{t}P_{1}=-\frac{5}{3}P_{0}\left({\nabla}\cdot\textbf{v}_{1}\right)-\frac{\alpha_{\parallel}cT_{0}}{6\pi}{\nabla}\cdot\left({\nabla}\times \textbf{B}_{1}\right)_{\parallel}+\\
-\frac{\aleph cT_{0}}{6\pi}\left[\textbf{B}_{0}\cdot\triangle\textbf{B}_{1}-{\nabla}\cdot\left(\textbf{B}_{0}\cdot{\nabla}\right)\textbf{B}_{1}\right]+\\
+\frac{2\chi_{\parallel}}{3}{\nabla}\cdot\left({\nabla}T_{1}\right)_{\parallel}+\frac{2\chi_{\perp}}{3}{\nabla}\cdot\left({\nabla}T_{1}\right)_{\perp}\;,\label{eq:P-1}
\end{multline}
and, assuming $(P_{0}+P_{1})=(n_{0}+n_{1})(T_{0}+T_{1})$ with \eref{eq:Continuità}, we easily obtain
\begin{equation}
\partial_{t}(P_{1}/P_{0})=-\nabla\cdot\textbf{v}_{1}+\partial_{t}(T_{1}/T_{0})\;
\label{eq:P-T}
\end{equation}
which can be used to get an equation for the solely variable $T_1$.

\section{Thermo-Galvanometric Instabilities}
\label{sec:2}

In the following, the linear perturbations are assumed to be local Fourier disturbances $\exp[i(\textbf{k}\cdot\textbf{r}-\omega t)]$, in which $\textbf{k}$ is the wave vector, $\textbf{r}$ the position vector and $\omega$ the angular frequency. For now, we do not consider dissipative effects such as viscosity, resistivity or thermal conductivity, and we treat three distinct regimes facing, firstly, only the presence of the Nernst coefficient (the analysis is performed, just for this case, separately for axial and radial perturbations to analytically outline the specific contribution of each case), secondly the Hall effects and then the global collisional contribution in the meridian plane.

\subsection{Nernst instability}

In this specific case, the following equations picture an ideal plasma in which the only collisional contribution comes from the Nernst effect. Furthermore, accordingly to the local approximation, we will consider $k\rb\gg1$. If we focus our attention on axial perturbations only, \emph{i.e.}, $\textbf{k}=k_{z}\textbf{e}_{z}$, it can be shown that the normal modes are characterized by the following dispersion relation where the Nernst contribution cancels out:
\begin{equation}\label{MRIdisprel}
\omega^{4}-\omega^{2}\left[\omega_{k}^{2}+2(\textbf{k}\cdot\textbf{v}_{A})^{2}\right]+(\textbf{k}\cdot\textbf{v}_{A})^{2}
\left[(\textbf{k}\cdot\textbf{v}_{A})^{2}-3\omega_{k}^{2}\right]=0\;.
\end{equation}
Setting $\omega_A=\textbf{k}\cdot\textbf{v}_{A}$, then instability can occur only if the condition
\begin{equation}
\omega_{A}^{2}<3\omega_{k}^{2}
\end{equation}
is verified. Thus, for large wave-length unstable modes surely appear, as in the standard MRI \citep{BH91}.

This kind of instability is not present if we choose radially oriented perturbations only, \emph{i.e.}, $\textbf{k}=k_{r}\textbf{e}_{r}$. To overcome the difficulties of finding an effective source of turbulence active in the region of low values of the $\beta$ parameter (\emph{i.e.}, $\beta\lesssim1$), in \citep{MNRAS} it has been shown how the presence of a non-zero Nernst coefficient causes a TMI, relevant in the parameter range where MRI is suppressed. In fact, by introducing a \emph{thermal frequency} $\omega_{T}=-c\aleph T_{0}k_{r}^{2}$, it is found that the ensuing relation, valid for the perturbed radial shifts of the fluid (defined by $v_{1r}=\p_t\xi_{r}$),
\begin{equation}
\omega^{4}\!+\!\Big(\frac{10\omega_{T}\Omega_{T}^{3}}{9\omega_{s}^{2}}-\Omega^{2}\Big)\omega^{2}+
i\frac{4}{3}\Omega_{T}^{3}\omega-\frac{10\Omega_{T}^{3}\omega_{T}}{9\omega_{s}^{2}}\Big(\omega_{k}^{2}+\frac{3}{5}\omega_{s}^{2}\Big)=0\;\label{eq:nernstkr}
\end{equation}
(here $\Omega_{T}^{3}=\omega_{T}\omega_{A(r)}^{2}$, $\omega_{s}=k_{r}v_{s}$ and $\Omega^{2}=\omega_{k}^{2}+\omega_{s}^{2}+\omega_{A(r)}^{2}$, with $\omega_{A(r)}=k_r v_A$) is substantially different from the corresponding MRI condition, and actually holds true even for supra-thermal fields. The detailed behavior of the TMI instability in function of the $\beta$ parameter is discussed in Sec.\ref{sec:3}.

The most interesting feature of these instabilities is the different response of the system once having fixed the direction of the perturbation wave-vector. One could regard this as a higher or lower coupling with the magnetic tension and pressure components of the plasma, making the two cases somewhat complementary: MRI vanishes in the radial wave-vector case, while TMI is not present for axially propagating perturbations. In principle, then, an intermediate case could be expected for poloidal perturbations of the fluid, in which the two instabilities coexist or prevail on each other at their respective typical scales. As we will see, this is not entirely true for the small wave length limit.

\subsection{Hall instability}

Let us for now drop the results of TMI and focus our attention to the case in which only the Hall effect be present. In this case, it is convenient to introduce the \emph{Hall frequency} $\omega_{\Re}=B_{0}/c^{2}\rho_0\left|\Re\right|$. It is a straightforward matter to show that this quantity is the ion Larmor frequency of the plasma and is only present as a ``scaling factor'': in other words, the Hall effect influence on the plasma configuration depends on the ratio of the gyration period of ions and the characteristic timescales relative to the plasma, which in general is small.

The dispersion relation is obtained through \erefs{eq:congelamento completo} and \reff{eq:P-T}, written in their scalar form, and read as
\begin{multline}
x+x\frac{\omega_{k}^{2}}{\omega_{\Re}^{2}}+\frac{5x-3\zeta_{k}}{2}\frac{\omega_{k}}{\omega_{\Re}}-3\zeta_{k}+\\
+\Big[(1+x+\zeta_{k})+x\frac{\omega_{A}^{2}}{\omega_{\Re}^{2}}-\frac{3}{2}\frac{\omega_{k}}{\omega_{\Re}}\Big]y^{2}+y^{4}=0\;,\label{eq:dispersionehallkrkzadim}
\end{multline}
where we have introduced the following dimensionless quantities: $y=-i\omega/\omega_{A}$, $x=k^{2}/k_{z}^{2}$, and $\zeta_{k}=\omega_{k}/\omega_{A}$. It is immediate to verify that the stability condition of the system stands as
\begin{equation}
\omega_{A}^{2}>3\omega_{k}^{2}/x(1+2\omega_{k}/\omega_{\Re})\;.\label{eq:condinsthallkrkz}
\end{equation}
This is a more constraining condition with respect to the MRI one, which is now slightly suppressed. The presence of the Hall coefficient alone does not, however, change the physics of the chosen configuration, and ends up constituting a small correction of its Magneto-rotational counterpart.

\subsection{Collision-driven instability}\label{subsec:itmi}

Let us now face the coupling, of collisional nature, between the magnetic field and the temperature gradient of the disk (we specify that MRI is instead based on the coupling between magnetic field and angular velocity gradient). In principle, one could consider the divergence of \eref{eq:Conservazioneqdmvettoriale} (together with \erefs{eq:congelamento completo}, \reff{eq:P-1} and \reff{eq:P-T}), to obtain the corresponding dispersion relation for the system. To shorten up the calculations, we here consider \erefs{eq:congelamento completo} and \reff{eq:Conservazioneqdmvettoriale} in their scalar form (we remark that we are assuming an incompressible fluid such that $\nabla\cdot\textbf{v}=0$):
\begin{subequations}
\begin{align}
\p_{t}v_{1r}-2\omega_{k}v_{1\phi}=-\frac{3}{5}\frac{\omega_{s}^{2}}{k_{r}^{2}}\;\p_{r}\frac{P_{1}}{P_0}+\nn\qq\qq\qq\\
+\frac{\omega_{A}^{2}}{B_{0z}k_{z}^\textbf{}{2}}\left(\partial_{z}B_{1r}-\partial_{r}B_{1z}\right)\;,\label{eq:ur-3-1}\\
\p_{t}v_{1\phi}+\frac{\omega_{k}}{2}v_{1r}=\frac{\omega_{A}^{2}}{B_{0z}k_{z}^{2}}\p_{z}B_{1\phi}\;,\label{eq:uphi-1}\\
\p_{t}v_{1z}=-\frac{1}{\rho_0}\p_{z}\left(P_{1}+P_{1}^{m}\right)\;,\\
\p_{t}\frac{T_{1}}{T_0}=\p_{t}\frac{P_{1}}{P_0}=-\frac{10}{9}\frac{\omega_{T}\omega_{A}^{2}}{B_{0z}\omega_{S}^{2}k_{z}^{2}}\left(\partial_{r}\partial_{z}B_{1r}-\partial_{r}^{2}B_{1z}\right)+\q\nn\\
-\frac{4\omega_{\alpha}}{3\beta B_{0z}k_{z}^{2}}\;\partial_{r}\partial_{z}B_{1\phi}\;,\label{eq:P-1-1}
\end{align}
\end{subequations}
where $\omega_{\alpha}=\alpha_{\parallel}cT_0k_{z}^{2}/B_{0z}$ and $P_{1}^{m}=\textbf{B}_0\cdot\textbf{B}_1/4\pi$. Similarly, for the induction equation, we get
\begin{subequations}
\begin{align}
\p_{t}B_{1r}=B_{0z}\p_{z}v_{1r}-\frac{B_{0z}\omega_{T}}{k_{r}^{2}}\;\partial_{r}\partial_{z}\frac{T_{1}}{T_0}\;,\label{eq:br}\\
\p_{t}B_{1\phi}=B_{0z}\p_{z}v_{1\phi}-\frac{3}{2}\omega_{k}B_{1r}+
\frac{B_{0z}\omega_{\alpha}}{k_{z}^{2}}\;\p_{r}\p_{z}\frac{T_{1}}{T_{0}}\;,\label{eq:bphi}\\
\p_{t}B_{1z}=-B_{0z}\partial_{r}v_{1r}+\frac{B_{0z}\omega_{T}}{k_{r}^{2}}\;\p_{r}^{2}\frac{T_{1}}{T_{0}}\;.\label{eq:bz-2-1}
\end{align}
\end{subequations}
Deriving \eref{eq:ur-3-1} with respect to $t$ and $r$, \erefs{eq:P-1-1} and \reff{eq:bphi}
with respect to time, and using the expressions for $\partial_{t}B_{1r}$ and $\partial_{t}B_{1z}$, the following set of three coupled differential equations is obtained:
\begin{subequations}
\begin{align}
(\p_{t}^{2}+\omega_{k}^{2})\p_{r}v_{1r}-\frac{2\omega_{k}\omega_{A}^{2}}{B_{0z}k_{z}^{2}}\partial_{r}\partial_{z}B_{1\phi}+\qq\qq\qq\nn\\+
\frac{3}{5}\omega_{s}^{2}\left(\frac{\partial_{r}^{2}}{k_{r}^{2}}+\frac{3}{2}\frac{\partial_{t}}{\omega_{T}}\right)\partial_{t}\frac{T_{1}}{T_{0}}=0\;,\label{eq:equazione1 nernstkrkz}
\end{align}
\begin{align}
-2B_{0z}\omega_{k}\partial_{z}v_{1r}+\Big(\frac{\omega_{A}^{2}}{k_{z}^{2}}\partial_{z}^{2}-\partial_{t}^{2}\Big)B_{1\phi}+\nn\qq\qq\qq\\
+B_{0z}\Big(\frac{3}{2}\frac{\omega_{k}\omega_{T}}{k_{r}^{2}}+\frac{\omega_{\alpha}}{k_{z}^{2}}\partial_{t}\Big)\partial_{r}\partial_{z}\frac{T_{1}}{T_{0}}=0\;,\\
\frac{10}{9}\frac{\omega_{T}\omega_{A}^{2}}{\omega_{S}^{2}k_{z}^{2}}\partial_{r}(\p_r^2+\p_z^2) v_{1r}+\frac{4\omega_{\alpha}}{3\beta B_{0z}k_{z}^{2}}\partial_{r}\partial_{z}\partial_{t}B_{1\phi}+\nn\qq\\
+\Big(\partial_{t}^{2}-\frac{10}{9}\frac{\omega_{T}^{2}\omega_{A}^{2}}{\omega_{S}^{2}k_{r}^{2}k_{z}^{2}}\partial_{r}^{2}(\p_r^2+\p_z^2)\Big)\frac{T_{1}}{T_{0}}=0\;.
\end{align}
\end{subequations}
It is straightforward to show that the dispersion relation takes the form:
\begin{align}
i\omega^{5}+i\Big[\Big(\frac{10}{9}\frac{\omega_{T}^{2}}{\omega_{s}^{2}}x-x-1\Big)\omega_{A}^{2}-\omega_{k}^{2}-\frac{3\beta}{4}(x-1)\omega_{\alpha}^{2}\Big]\;\omega^{3}+\nonumber\\	-\Big[\frac{2}{3}x\omega_{A}^{2}\omega_{T}+\frac{3}{2}\omega_{k}\omega_{T}\omega_{\alpha}\Big(1-\frac{6\omega_{s}^{2}}{5\omega_{T}^{2}}\Big)\Big]\;\omega^{2}+\nonumber\\	+i\Big[x\omega_{A}^{4}-3\omega_{A}^{2}\omega_{k}^{2}-\frac{10}{9}x\frac{\omega_{A}^{4}\omega_{T}^{2}}{\omega_{s}^{2}}+\nn\qq\qq\qq\qq\q\\
-\frac{10}{9}x\frac{\omega_{A}^{2}\omega_{k}^{2}\omega_{T}^{2}}{\omega_{s}^{2}}+\frac{6}{5}\omega_{k}\omega_{s}^{2}\omega_{\alpha}+\frac{3\beta}{4}(x-1)\omega_{k}^{2}\omega_{\alpha}^{2}\Big]\;\omega+\nn\\	+\frac{2}{3}x\omega_{A}^{4}\omega_{T}\Big[1+\frac{5\beta}{2}(x-1)\frac{\omega_{k}\omega_{\alpha}}{\omega_{s}^{2}}\Big]-\frac{3}{2}\omega_{k}^{3}\omega_{T}\omega_{\alpha}=0\;,\label{eq:dispersionekrkzn}
\end{align}
where we recall that $x=k^{2}/k_z^{2}$.

By introducing $\zeta_{\alpha}=\omega_{\alpha}/\omega_{A}$ and $\zeta_{T}=2\omega_{T}/3\omega_{A}$, \eref{eq:dispersionekrkzn} can be recast in the following dimensionless form (we recall that $y=-i\omega/\omega_A$):
\begin{align}
y^{5}+\Big[1+\zeta_{k}^{2}+\Big(1-\frac{3\zeta_{T}^{2}}{\beta(x-1)}\Big)x\Big]\;y^{3}+\nonumber\\
+\Big[\frac{9}{4}-\frac{\beta(x-1)}{\zeta_{T}^{2}}\zeta_{T}\zeta_{\alpha}\zeta_{k}-x\zeta_{T}\Big]\;y^{2}+\nonumber\\
+\Big[\Big(1-\frac{3\zeta_{T}^{2}}{\beta(x-1)}(1+\zeta_{k}^{2})\Big)x-3\zeta_{k}^{2}+\nn\qq\qq\qq\qq\\+
\frac{6}{5}\beta(x-1)\zeta_{k}\zeta_{\alpha}(1+\frac{5}{8})\zeta_{k}\zeta_{\alpha}\Big]\;y+\nonumber\\
-x\zeta_{T}(1+3\zeta_{k}\zeta_{\alpha})-\frac{9}{4}\zeta_{k}^{3}\zeta_{T}\zeta_{\alpha}=\;0\;.\label{eq:adim}
\end{align}
This is a quintic equation in $y$ and, in general, it has no analytic solution. \emphr{In order to make a direct comparison with the MRI case, we study here pure real solutions only, \emph{i.e.}, we consider (absolute) non propagating collision-driven instabilities. A discussion on the convective contribution is provided in the next Subsection.}

\begin{figure}
\centering
\includegraphics[width=0.75\linewidth]{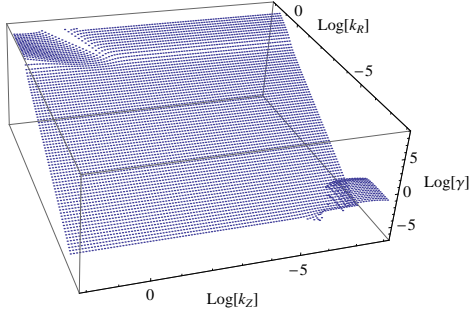}
\caption{Plot of the leading pure real solution of \eref{eq:adim}, namely $y=\gamma>0$, as a function of the components of the wave-vector $\mathbf{k}$, pictured in a logarithmic plot. In this case, we have set: $\beta=0.1$, $T=10^{4}$K, $n=10^{12}\,\textrm{cm}^{-3}$. Remember that the vertical scales are bound to be less than twice the value of the characteristic scale height, $H=\sqrt{2}u_{s}/\omega_{k}$, so that it must be $k_{z}<3\times10^{-6}$.}
\label{fig:1}
\end{figure}

\emphr{As expected, in the $\textbf{k}=k_{z}\textbf{e}_{z}$ and $\textbf{k}=k_{r}\textbf{e}_{r}$ cases, we can reach the above introduced relations for MRI (\eref{MRIdisprel}) and TMI (\eref{eq:nernstkr}), respectively. The overall instability curve is \emph{de facto} a superposition of these two modes: by their very own nature, the couplings of the total magnetic field with the angular velocity (generating the MRI) and with the thermal gradients (yielding the TMI) tend to generate strong instabilities in the fluid, but at different scales. In fact, while MRI dominates the large spatial scale, TMI is relevant when radial scales are kept sufficiently small. The response of the plasma to magneto-rotational and thermo-magnetic effects can thus be regarded as two separate, coexisting instabilities.} This is well described in \figref{fig:1}, where the (leading) instability curve has been depicted as a function of the components of the wave-vector $\mathbf{k}$. As for the role of the parallel term related to the electromotive force, its relevancy is limited to the small scale region of the perturbations. Its net effect is a relevant damping of the instability up to several order of magnitude compared to what the pure TMI would be. Note, however, that this behavior emerges at scales that tend to approach the Debye length of the plasma as the local temperature rises, so that it is important only for a narrow range of temperatures and densities. Moreover, as we will discuss in the next Section, the presence of resistive effects efficiently dampens the small scale modes of the plasma, so that this phenomenon, while worth mentioning, is not observable in actual plasma disks.

\subsection{\emphr{Discussion over convective instabilities}}
\begin{figure}
\centering
\includegraphics[width=0.7\linewidth]{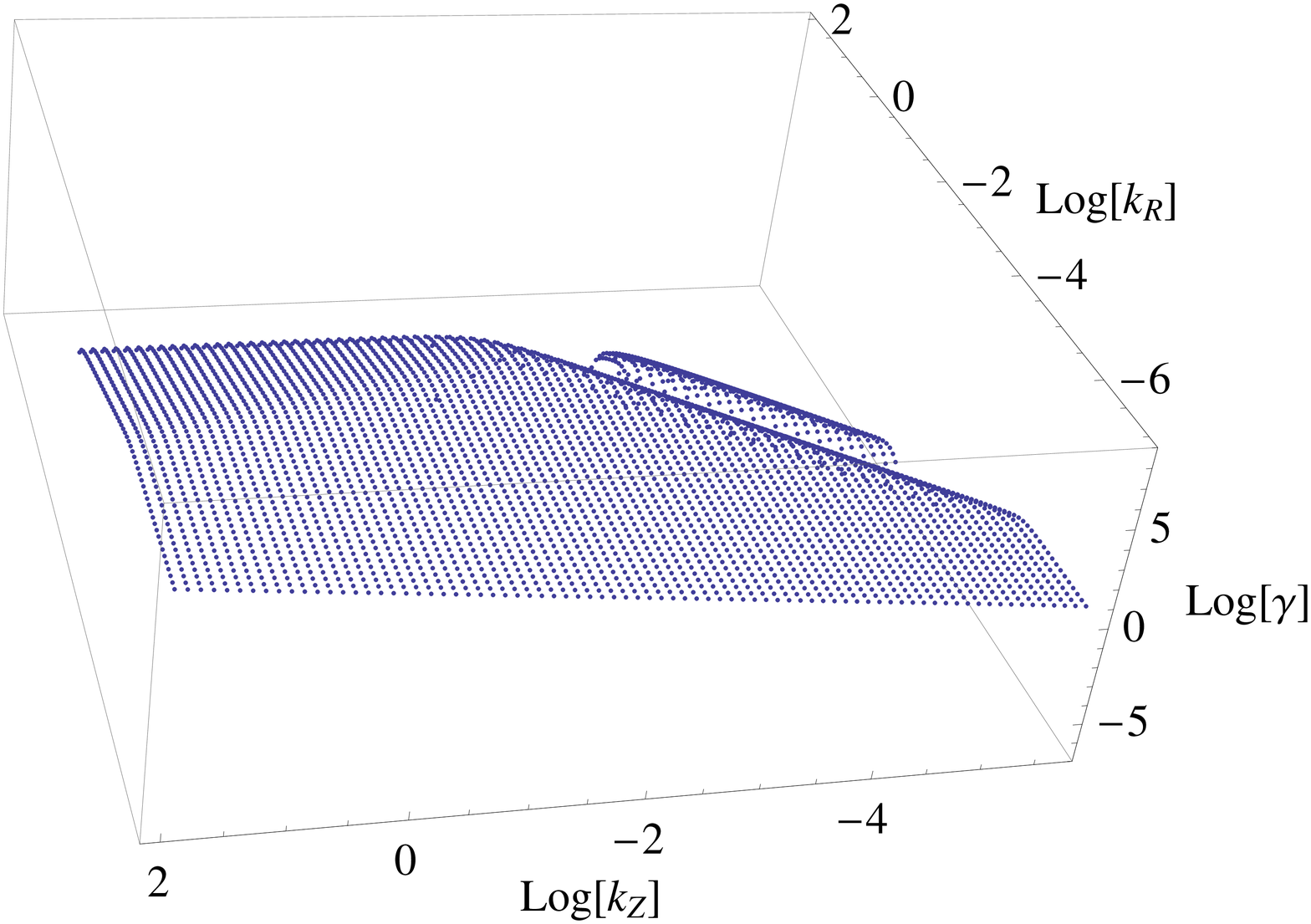}\\
\includegraphics[width=0.7\linewidth]{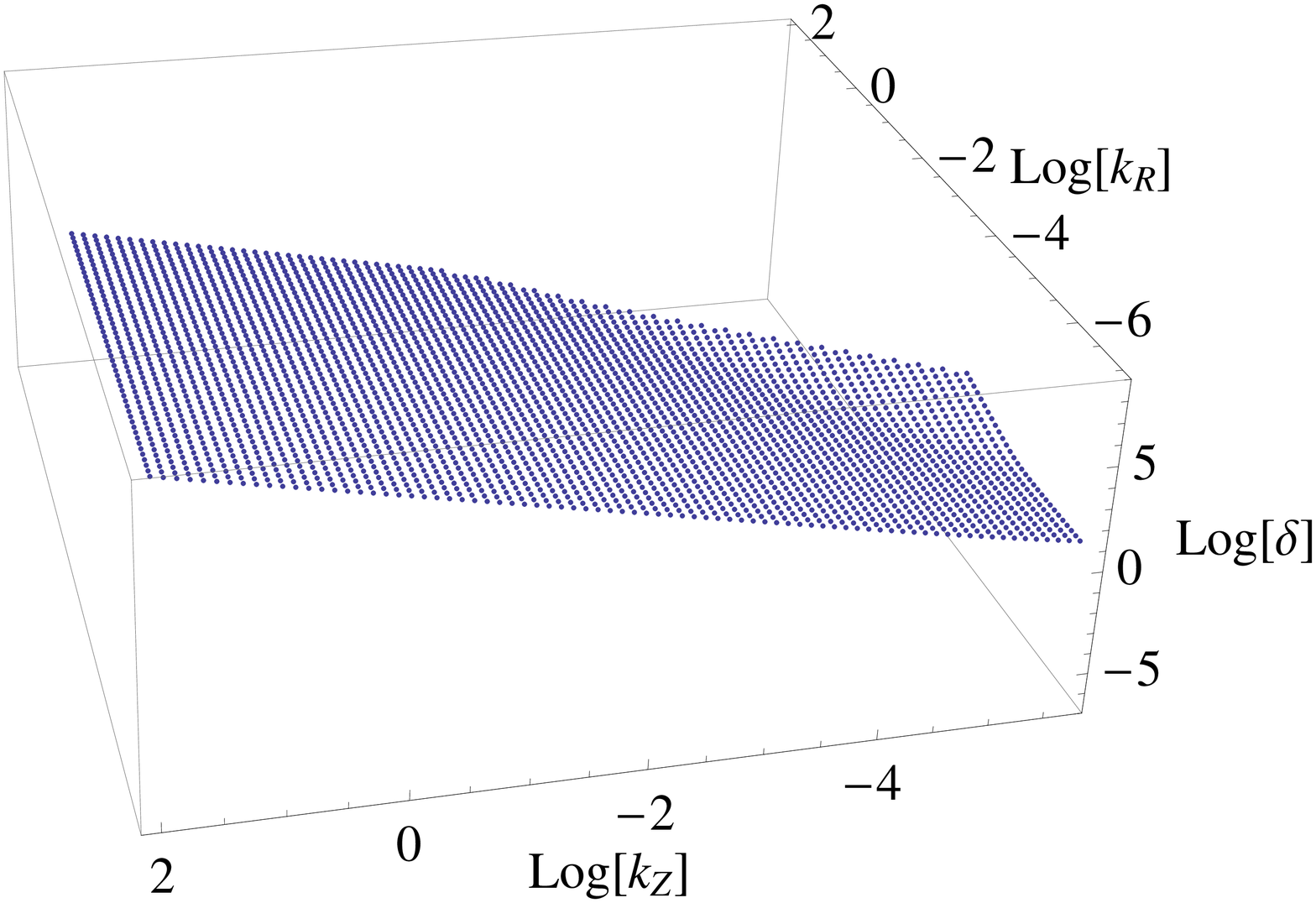}
\caption{Leading convective solutions of \eref{eq:adim}, namely $y=\gamma+i\delta$. The parameters are set as for \figref{fig:1}.}
\label{fig:2}
\end{figure}

Despite our analysis, as discussed above, is aimed to compare absolute instabilities with respect to MRI (which is indeed non-propagative), we now investigate the behavior of the analyzed instabilities when complex and conjugated solutions, $y=\gamma+i\delta$, of the dispersion relation \reff{eq:adim} are considered. The presence of a real frequency makes such instabilities convective in character and, in \figref{fig:2}, we plot the real (upper panel) imaginary  (lower panel) part of this kind of solutions.

It is immediate to recognize that, for large enough values of $k_r$, the contribution of the convective instability is negligible with respect to the corresponding absolute one, depicted in \figref{fig:1}. A new feature arises when $k_r$ is sufficiently small, \emph{i.e.}, at large spatial scales, where the convective instability is significantly dominant being of the same order of the MRI profile. Thus, the large scales of the plasma disk are characterized by an unstable convective mode whose growth rate increases monotonically with the perturbation wave-length. However, we note that the instability profile must be cut by the natural size of the disk, which fixed the minimum value of the admissible wave-number.

The behavior of the corresponding real part of the solution, \emph{i.e.}, the frequency of the propagating mode, outlines a linear behavior in the region where the convective instability lives. When we analyze the complex and conjugated solutions, the collision-driven instability acquires the peculiar property to be relevant at large scales too, becoming competitive with the MRI even in fixing the global properties of the disk. We conclude by stressing how the convective nature of the present instability is essentially due to the dependence on $k_z$ of the real part of $\omega$. In fact, while the real frequency is almost independent of $k_r$, it roughly outlines the linear behavior $\delta\simeq 10^{7}k_z$, \emph{i.e.}, the group speed is about $10^{2}$km/s. Thus, we can claim how the complex and conjugated solution is really associated to a convective instability when it has a non-zero vertical component of the wave-vector.

\section{Dissipative Effects}
\label{sec:3}

In general, dissipative terms tend to suppress the growth of unstable modes in a plasma. But exactly, in which way these effects are to be taken into account? Here, we will not address the problematics of handling a viscous fluid, rather we focus our attention to electrical and thermal conductivity. Following \citep{LP81}, the expressions of their parallel and orthogonal components are
\begin{align}
\sigma_{\parallel}&=\frac{4\sqrt{2}}{\pi^{3/2}}\frac{T^{3/2}}{e^{2}L_{e}m^{1/2}}\;,\qquad\quad
\sigma_{\perp}=\frac{3\pi^{1/2}}{\sqrt{2}}\frac{e^{2}n}{m\nu_{ei}}\;,\\\nn
\chi_{\parallel}&=\frac{16\sqrt{2}}{\pi^{3/2}}\frac{T^{5/2}}{e^{4}L_{e}m^{1/2}}\;,\qquad\;\;
\chi_{\perp}=\begin{cases}
\frac{2}{3\pi^{1/2}}\frac{nT\nu_{ii}}{m_{i}\omega_{Bi}^{2}} & \omega_{Bi}\gg\nu_{ii}\;,\\
\frac{17}{6\pi^{1/2}}\frac{nT\nu_{ee}}{m\omega_{Be}^{2}} & \omega_{Bi}\ll\nu_{ii}\;,
\end{cases}
\end{align}
where, $m_i$ is the ion mass, $\nu_{ii}$ is the ion-ion collisional frequency and $\omega_{Bi}=eB/cm_{i}$.

Let us now run a dimensional analysis of the equations in order to assess the actual relevance of each dissipative term. We remark that the parallel component of the thermal conductivity $\chi_{\parallel}$ results relevant for every case of astrophysical interest, since it can be neglected only for
\begin{equation}
\frac{T^{4}}{n\sqrt{\beta}}\ll\frac{3\pi}{32}e^{6}mc^{2}L_{e}^{2}\;,
\end{equation}
which, however, violates the requirement $\omega_{Be}\gg\nu_{ee}$. Hence, we will always consider its contribution in the following calculations. On a closer look, the perturbed form of \erefs{eq:calore} and \reff{eq:ohmgeneralizzata} gives us an insight on which of the remaining terms play an important role in the onset of unstable modes. In particular, the $\beta$ values determine which resistive effect has to be taken into account: in the $\beta\gg1$ limit, the transverse thermal conductivity is the only additional relevant effect; on the other hand, if $\beta\ll1$, one has to consider both of the coefficients associated to the electrical conductivity, along with $\chi_{\perp}$ if the background magnetic field is strong enough to verify $\omega_{Bi}\gg\nu_{ii}$. 

Following the same procedure of the previous Section, it is found that the full dispersion relation for the two complementary cases described above can be obtained by considering the following system of differential equations (we use $\BB=(\p_z B_{1r}-\p_r B_{1z})$):
\begin{subequations}
\begin{multline}
(\p_{t}^{2}+\omega_{k}^{2})v_{1r}-\frac{\omega_{A}^{2}}{B_{0z}k_{z}^{2}}\p_{t}\BB+\\	-\frac{2\omega_{k}\omega_{A}^{2}}{Bk_{z}^{2}}\partial_{z}B_{1\phi}+\frac{3\omega_{s}^{2}}{5k_{r}^{2}}\partial_{r}\partial_{t}\frac{T_{1}}{T_{0}}=0\;,
\end{multline}\vspace{-6mm}
\begin{multline}
-2B_{0z}\omega_{k}\partial_{z}v_{1r}-\Big(\frac{\omega_{A}^{2}}{\omega_{\Re}k_{z}^{2}}\partial_{z}\partial_{t}+\frac{3\omega_{k}\oso}{2k_{z}^{2}}\partial_{z}\Big)\BB+\\
-\Big(\partial_{t}^{2}-\frac{\omega_{A}^{2}}{k_{z}^{2}}\partial_{z}^{2}-\frac{\omega_{\sigma_{\parallel}}}{k_{z}^{2}}\partial_{r}^{2}\partial_{t}-\frac{\oso}{k_{r}^{2}}\partial_{z}^{2}\partial_{t}+\frac{3\omega_{k}\omega_{A}^{2}}{2\omega_{\Re}k_{z}^{2}}\partial_{z}^{2}\Big)B_{1\phi}+\\
+\Big(\frac{3B_{0z}\omega_{k}\omega_{T}}{2k_{r}^{2}}\partial_{r}\partial_{z}+\frac{B_{0z}\omega_{\alpha}}{k_{z}^{2}}\partial_{r}\partial_{z}\partial_{t}\Big)\frac{T_{1}}{T_{0}}=0\;,
\end{multline}\vspace{-6mm}
\begin{multline}
B_{0z}(\p_r^2+\p_z^2)v_{1r}+\Big(-\partial_{t}+\frac{\omega_{\sigma_{\perp}}}{k_{z}^{2}}(\p_r^2+\p_z^2)\Big)\BB+\nn\\+\frac{\omega_{A}^{2}}{\omega_{\Re}k_{z}^{2}}\partial_{z}(\p_r^2+\p_z^2)B_{1\phi}-\frac{B_{0z}\omega_{T}}{k_{r}^{2}}\partial_{r}(\p_r^2+\p_z^2)\frac{T_{1}}{T_{0}}=0\;,
\end{multline}\vspace{-6mm}
\begin{multline}
	\frac{10\omega_{T}\omega_{A}^{2}}{9B_{0z}\omega_{s}^{2}k_{z}^{2}}\partial_{r}\BB+\frac{4\omega_{\alpha}}{3\beta B_{0z}k_{z}^{2}}\partial_{r}\partial_{z}B_{1\phi}+\\
+\Big(\partial_{t}-\frac{\omega_{\chi_{\parallel}}}{k_{z}^{2}}\partial_{z}^{2}-\frac{\omega_{\chi_{\perp}}}{k_{r}^{2}}\partial_{r}^{2}\Big)\frac{T_{1}}{T_{0}}=0\;,
\end{multline}
\end{subequations}
where $\omega_{\sigma_{\parallel,\perp}}=c^2k_{r,z}^{2}/4\pi\sigma_{\parallel,\perp}$, and $\omega_{\chi_{\parallel,\perp}}=2\chi_{\parallel,\perp}k_{z,r}^{2}/3n$.

\begin{figure}
\includegraphics[width=0.47\linewidth]{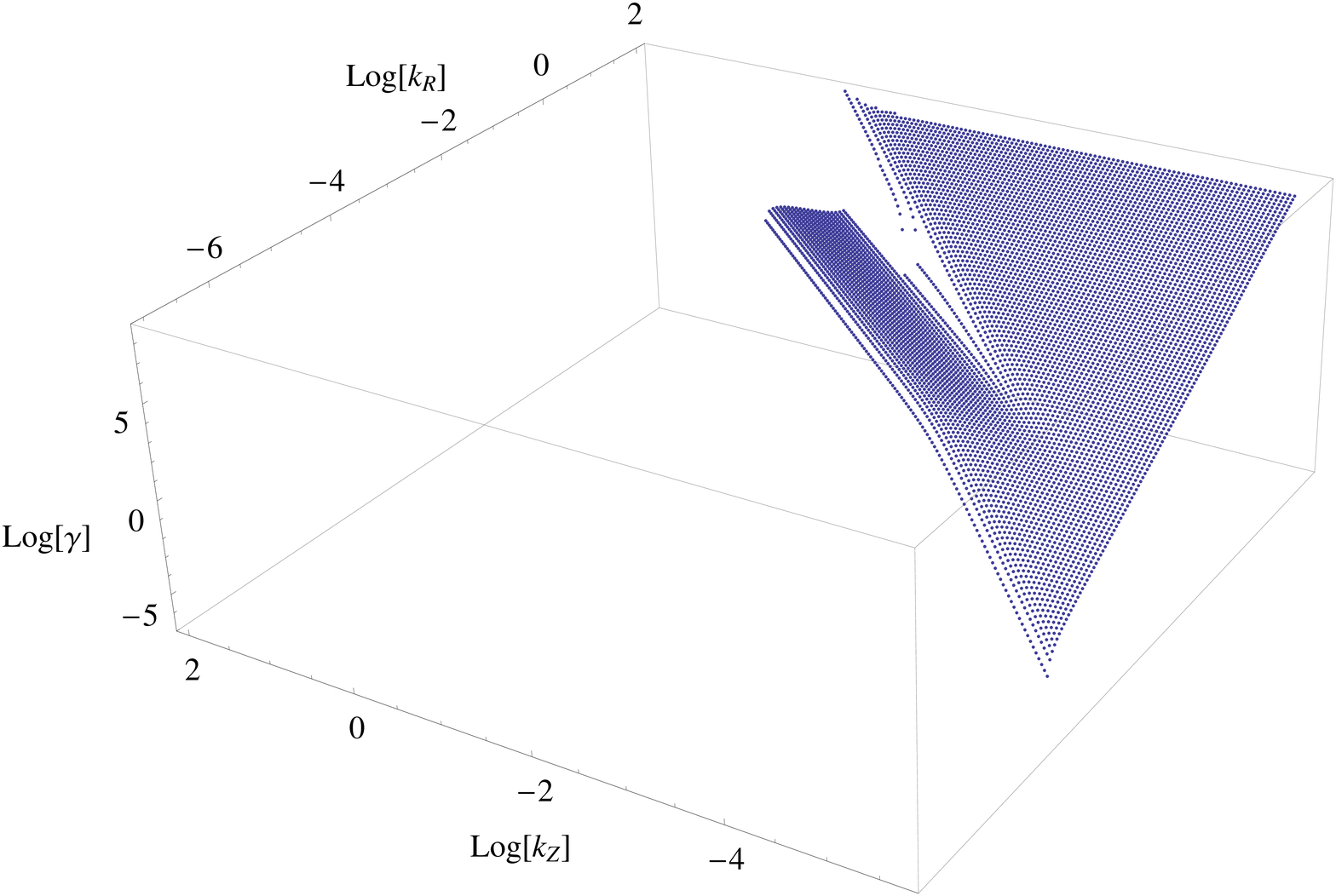}
\includegraphics[width=0.47\linewidth]{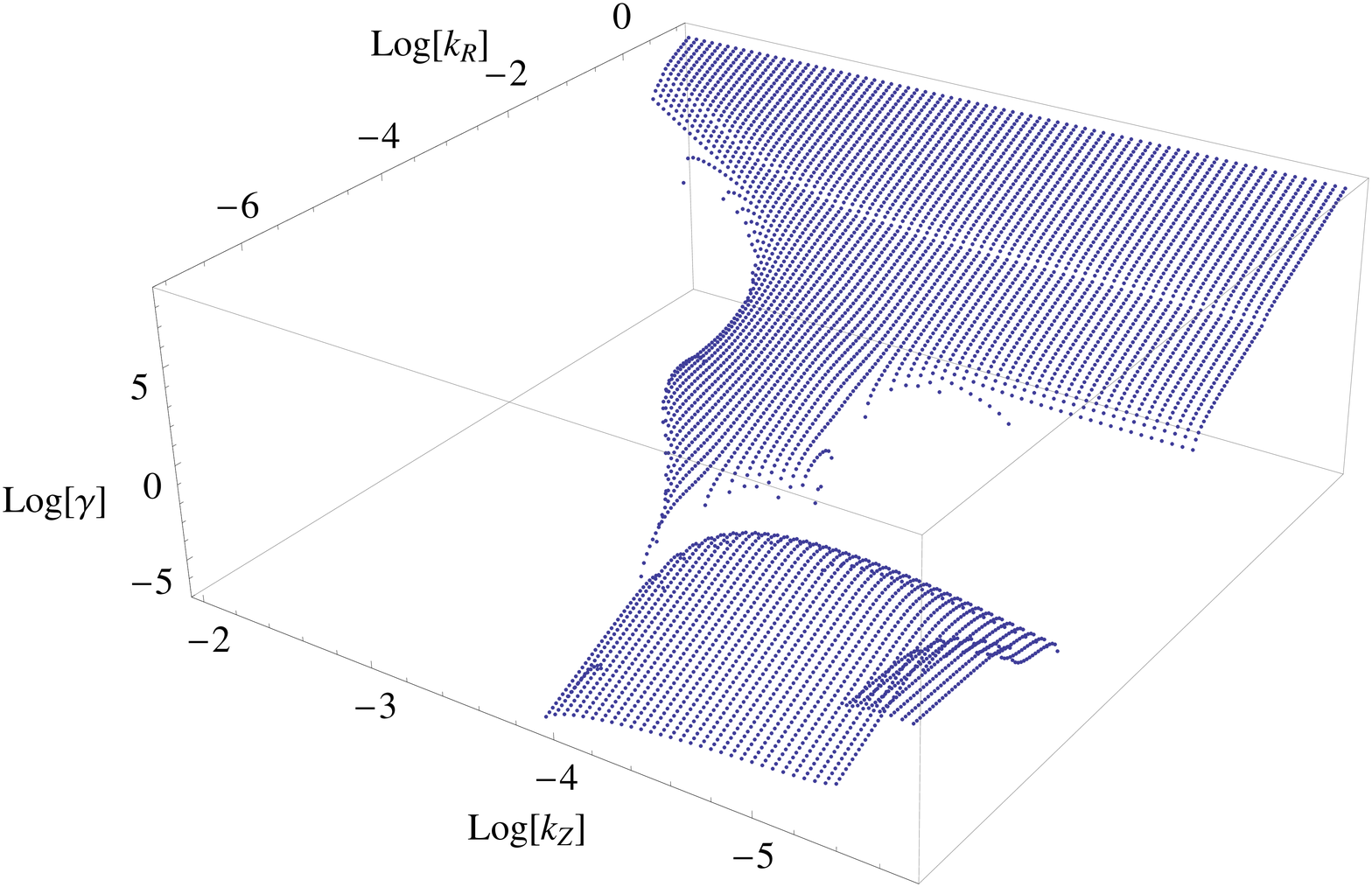}
\caption{Left panel - Low-$\beta$ limit: Log-Log plot of $\gamma$ as a function $k_{r}$ and $k_{z}$. The plasma parameters are set as $T=10^{4}\textrm{K}$, $n=10^{12}\textrm{cm}^{-3}$, $\beta=0.1$ and $\omega_{k}=1\textrm{s}^{-1}$. Note how resistive terms efficiently suppress the instability for all the small vertical scales verifying $k_{z}\gg k_{r}$. Right panel - High-$\beta$ limit: plot of $\gamma$ for a plasma with $T=10^{4}\textrm{K}$, $n=10^{10}\textrm{cm}^{-3}$, $\beta=10$ and $\omega_{k}=1\textrm{s}^{-1}$. TMI is here strongly suppressed and exists only at very small radial scales of order $k_r^{-1}<10\textrm{cm}$. The MRI branch of the curve is only slightly damped, and retains all the peculiarities observed in the ideal case.\protect\label{fig:3}}
\end{figure}

The resulting dispersion relation is rather intricate and very little can be inferred from its form, which thus we do not present here for the sake of brevity. We can however get a glimpse of the overall behavior of the plasma just looking at \figref{fig:3}, in which the numerically integrated unstable \emphr{(both absolute and convective)} solution $\gamma(k_{r},\,k_{z})$ of the dispersion relation is depicted, in the two limits of high and low $\beta$ and for a background magnetic field strength small enough in order to obtain $\omega_{Bi}<\nu_{ii}$ (\emph{i.e.}, for which the only contribution of the transverse thermal conductivity is electronic). If this last condition were not imposed, the system would be stable to perturbation of thermo-magnetic nature, and only the coupling between $B_{0z}$ and the angular velocity gradient could generate instability. Hence, in this case, only the MRI is present. In general, it is the very longitudinal component of the thermal conductivity which efficiently suppresses the instability for small axial scales of the perturbations and, in that region, the system is now stable regardless of the presence of thermo-galvanometric phenomena. Thus, it tends to stabilize the plasma for high values of the wave vector, in the region in which the condition $k_{z}\gg\,k_{r}$ holds true. 

For what concerns the role of the other dissipative components, by comparing the low-$\beta$ curve of the solutions with respect to the high-$\beta$ one, it appears that electric resistivity has a much less pronounced damping effect than the orthogonal thermal coefficient $\chi_{\perp}$, at least in the small- and meso-spatial scale regions. Nevertheless, thermo-magnetic and magneto-rotational unstable modes of the plasma still survive for a wide range of choices for the wave vector, especially in the $\beta\ll1$ limit. \emphr{It is worth noting how, as shown in \figref{fig:3}, in the present dissipative case, the propagating (convective) instabilities do not provide significant information with respect to the absolute case, in the sense that their growth rates are comparable.} Hence, the TMI can indeed be regarded as another possible candidate to the disk overall instability.

\section{Conclusions}
Over forty years of studies have been focused on the understanding of the many facets of accretion disk energy outflow. The core of this phenomenon lies in turbulent behavior and it is now clear that the presence of background magnetic fields plays an active role in triggering turbulence-enhanced transport via linear instabilities in the plasma \citep{mri}. 

We add a contribution to this issue by showing that the inclusion of thermo-magnetic and galvanometric effects in the system favor the instability profile of accretion disks. Moreover, we outline how the contribution of electric and thermal conductivities generate a damping of the overall unstable mode, suppressing it for several choices of the wave vector. Still, TMI is nonetheless present to some extent, especially in the low-$\beta$ region, for sufficiently small values of the background magnetic field, and actively contributes to the disk overall instability.

\emphr{In the present study, we analyze the collision-driven unstable modes in the case of an exactly vertical magnetic field. Such a choice can appear as a simplification hypothesis, but it is actually a reliable model for a thin disk profile. In fact, the structure of a compact star magnetic field is essentially poloidal (the toroidal component almost identically vanishes) and it closely resembles a dipole-like morphology. Therefore, for the case of a thin disk mainly distributed around the equatorial plane, the assumption of a vertical background magnetic field provides a very reliable model of the real external magnetic profile in which the plasma disk is embedded. However, it is worth noting that, across the disk, a toroidal magnetic field is generated by a dynamo effect \citep{BS05,Sc04}. This consideration suggests that the extension of the present work to the case in which the magnetic field has a toroidal component too, is a natural perspective for an astrophysical implementation of the present instabilities. Nonetheless, we stress that a significant enhancement of the toroidal magnetic field is ensured by turbulence effects, see for instance the $\alpha$-$\omega$ effect \citep{SKR66,K80}, which are triggered by the instability profiles here discussed. Thus, when studying the onset of the plasma disk turbulence via linear unstable modes, it is a reliable scenario to remove the toroidal component of the magnetic field, which, if present, is still not relevant enough.}

\emphr{Although it is rather difficult to get a detailed information on the accreting plasma disk from the observations, however studies exist which outline inconsistencies of the standard paradigm for accretion, see for instance \citep{Li10}. For what concerns the present study and the competitive role of TMI with respect to MRI, we observe that the most promising astrophysical systems, in which our predictions can have a significant impact, are those ones characterized by low values of the plasma $\beta$ parameter and high accretion rates. In fact, for such a scenario, MRI can be significantly suppressed and the standard accretion paradigm meets difficulties in justifying the dissipation effects required by the observed accretion rates: in this respect, the most promising stellar systems, where the role of TMI can be expected to be relevant, are very cold X-ray binaries \citep{Ga10}.}

\emphr{Finally, we observe that the role played by TMI in astrophysics must rely on non-trivial correlations between different scales, so that phenomena occurring at micro-scales could realistically have an effect on meso-scales, as observed in laboratory plasma \citep{zonca}.}

\section*{Acknowledgments}
This work was partially developed within the framework of the \emph{CGW Collaboration}
(www.cgwcollaboration.it).

\end{document}